\documentclass[aps,prl,twocolumn,showpacs,superscriptaddress]{revtex4}

\usepackage{amssymb}
\usepackage{amsmath}
\usepackage[british]{babel}
\usepackage{graphicx}
\usepackage{color}

\begin{document}

\title{Anomalous diffusion and ergodicity breaking in heterogeneous diffusion
processes}

\author{Andrey G. Cherstvy}
\affiliation{Institute for Physics \& Astronomy, University of Potsdam,
14476 Potsdam-Golm, Germany}
\email{a.cherstvy@gmail.com}
\author{Aleksei V. Chechkin}
\affiliation{Akhiezer Institute for Theoretical Physics, Kharkov Institute
of Physics and Technology, Kharkov 61108, Ukraine}
\affiliation{Max-Planck Institute for the Physics of Complex Systems,
N{\"o}thnitzer Stra{\ss}e 38, 01187 Dresden, Germany}
\author{Ralf Metzler}
\affiliation{Institute for Physics \& Astronomy, University of Potsdam,
14476 Potsdam-Golm, Germany}
\affiliation{Department of Physics, Tampere University of Technology, 33101
Tampere, Finland}
\email{rmetzler@uni-potsdam.de}

\date{\today}

\begin{abstract}
We demonstrate the non-ergodicity of a simple Markovian stochastic processes
with space-dependent diffusion coefficient $D(x)$. For power-law forms $D(x)
\simeq|x|^{\alpha}$, this process yield anomalous diffusion of the form
$\langle x^2(t)\rangle\simeq t^{2/(2-\alpha)}$. Interestingly, in both the sub-
and superdiffusive regimes we observe weak ergodicity breaking: the scaling of
the time averaged mean squared displacement $\overline{\delta^2}$ remains
\emph{linear\/} and thus differs from the corresponding ensemble average
$\langle x^2(t)\rangle$. We analyze the non-ergodic behavior of this process in
terms of the ergodicity breaking parameters and the distribution of amplitude
scatter of $\overline{\delta^2}$. This model represents an alternative
approach to non-ergodic, anomalous diffusion that might be particularly
relevant for diffusion in heterogeneous media.
\end{abstract}


\maketitle

Since the early systematic studies of Perrin and Nordlund \cite{perrin}, single
particle tracking has become a routine method to trace individual particles in
microscopic systems \cite{spt}, but also of animal \cite{ran} and human
\cite{dirk} motion patterns. In a wide range of systems and over many time and
length scales these measurements reveal anomalous diffusion with mean squared
displacement (MSD) $\langle x^2(t)\rangle\simeq t^p$ \cite{report}. Subdiffusion
($0<p<1$) was found for the motion of tracers in living biological cells
\cite{weber,krapf,lene,seisenhuber,elbaum} or of bacteria in biofilms
\cite{rogers}. Superdiffusion ($p>1$) is observed for active motion in
biological cells \cite{elbaum}, of mussels \cite{mussels}, plant lice
\cite{aphids}, or higher animals and humans \cite{ran,dirk}.

Of particular current interest are the ergodic properties of a stochastic
process: is the information obtained from time averages of a single or few
trajectories representative for the ensemble \cite{pt,igor}? Of the
various stochastic models for anomalous diffusion, diffusion on random
fractals has been shown to be ergodic \cite{yazmin}. Fractional Brownian
and fractional Langevin equation motion driven by long-range correlated
Gaussian noise are transiently non-ergodic \cite{deng,jae,goychuk}.
In contrast, subdiffusive Scher-Montroll continuous time random walks (CTRW)
\cite{montroll} with diverging characteristic waiting time exhibit weak
ergodicity breaking \cite{web}, and time averages remain random even in the
long time limit \cite{he}. Weak ergodicity breaking was indeed observed
for the blinking dynamics of quantum dots \cite{eli}, protein motion in human
cell walls \cite{krapf}, and lipid granule diffusion in yeast cells \cite{lene}.

Can we come up with a simple stochastic model that still features such intricate
non-ergodic behavior? Here we show that the MSD of heterogeneous diffusion
processes (HDPs) with space dependent diffusion constant $D(x)\simeq|x|^{\alpha}$
scales like $\langle x^2(t)\rangle\simeq t^p$ with $p=2/(2-\alpha)$, while the
time averaged MSD $\overline{\delta^2}$ scales linearly in both sub- ($\alpha
<0$) and superdiffusive ($\alpha>0$) regimes. We quantify the non-ergodicity
in terms of the ergodicity breaking parameters and the amplitude scatter
distribution of $\overline{\delta^2}$. We showcase the analogies and
differences between HDPs and other non-ergodic diffusion processes.

Physically, a space dependent diffusivity appears a natural description for
diffusion in heterogeneous systems. Examples include Richardson diffusion in
turbulence \cite{richardson} as well as mesoscopic approaches to transport in
heterogenous porous media \cite{dentz} and on random fractals \cite{loverdo}.
Recently, maps of the local cytoplasmic diffusivities in bacterial and
eukaryotic cells showd a highly heterogeneous landscape \cite{kuehn}, recalling
the strongly time-varying diffusion coefficients of tracers in cells, see, e.g.,
Ref.~\cite{platani}.

\emph{Model.} We start with the Markovian Langevin equation for the displacement
$x(t)$ of a test particle in a heterogeneous medium with space dependent
diffusivity $D(x)$, namely, $dx/dt=\sqrt{2D(x)}\zeta(t)$.
Here, $\zeta(t)$ is white Gaussian noise with $\delta$-correlation $\langle
\zeta(t)\zeta(t')\rangle=\delta(t-t')$ and zero mean $\langle\zeta(t)\rangle=0$.
In the following we interpret the Langevin equation in the Stratonovich sense
\cite{risken}, to ensure the correct limiting transition for the noise with
infinitely short correlation times \cite{REM}. Specifically, for $D(x)$ we
employ the power-law form \cite{footnote2}
\begin{equation}
\label{diff-coeff-power}
D(x)=D_0|x|^{\alpha}.
\end{equation}
The dimension of $D_0$ is $[D_0]=\mathrm{cm}^{2-\alpha}\mathrm{sec}^{-1}$. In the
Stratonovich interpretation, we introduce the substitution $y=\int^x[2D(x')]^{
-1/2}dx'$, where $y(t)$ is the Wiener process \cite{risken}. For the initial
condition $P(x,0)=\delta(x)$ a compressed ($\alpha<0$) or stretched ($\alpha>0$)
Gaussian PDF
\begin{equation}
\label{PDF-HDP}
P(x,t)=\frac{|x|^{1/p-1}}{\sqrt{4\pi D_0t}}\exp\left(-\frac{|x|^{2/p}}{
(2/p)^{2}D_0t}\right)
\end{equation}
emerges (see Supplementary Material (SM) \cite{supp}), with $p=2/(2-\alpha)$.
The ensemble averaged MSD $\langle x^{2}(t) \rangle=\int x^2P(x,t)dx$ becomes
\begin{equation}
\label{MSD-theory}
\langle x^{2}(t)\rangle=\frac{\Gamma(p+1/2)}{\pi^{1/2}}\left(\frac{2}{p}\right)
^{2p}(D_0t)^p
\end{equation}
Thus, for $\alpha<0$ we find subdiffusion, and superdiffusion
for $\alpha>0$. Brownian motion with $\langle x^2(t)\rangle=2D_0t$ emerges for
$\alpha=0$, and $\alpha=1$ produces ballistic motion. The diffusion becomes
increasingly fast when $\alpha$ tends to 2.

Fig.~\ref{fig1} compares the PDF (\ref{PDF-HDP}) to simulations results
(see below). Compared to the Gaussian shape of Brownian motion ($\alpha=0$),
we observe a depletion in regions of higher diffusivity. For subdiffusion,
this causes the central dip of the PDF, while for superdiffusion probability
is shifted towards the origin. Note that the shape of the PDFs is significantly
different from those of CTRW processes. In particular, for the subdiffusive
case, the CTRW-PDF has a pronounced cusp at the origin \cite{report}.
Curiously, the subdiffusive shape in Fig.~\ref{fig1} resembles the propagator
for retarded wave motion \cite{abm}.

\begin{figure}
\includegraphics[width=8.8cm]{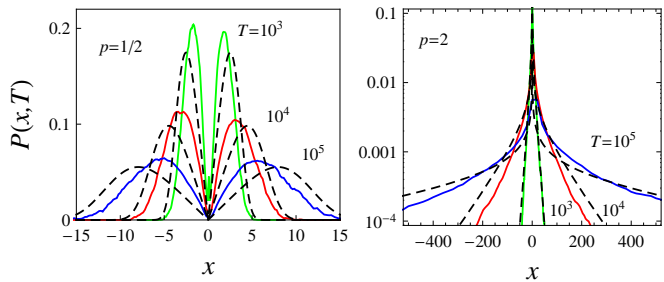}
\caption{Time evolution of the PDF for sub- (Left) and superdiffusion (Right).
We compare simulations results (full lines) to the analytical result (dashed)
of Eq.~(\ref{PDF-HDP}). The small discrepancy is due to the rectified form of
the diffusion coefficient at the origin implemented in simulations (see text).}
\label{fig1}
\end{figure}

To characterize the HDP, we calculate the position autocorrelation (see SM
\cite{supp}) for the case $t_2>t_1$,
\begin{eqnarray}
\nonumber
\langle x(t_1)x(t_2)\rangle&=&\frac{2^{p+1}}{\sqrt{\pi}}\frac{\Gamma(p+1)
\Gamma(\frac{p}{2}+1)}{p^{2p}\Gamma(\frac{p}{2}+\frac{1}{2})}[D_0t_1]^{(p+1)/2}\\
&&\hspace*{-2.6cm}
\times[D_0(t_2-t_1)]^{(p-1)/2}\,_2F_1\left(\frac{1-p}{2},\frac{p}{2}+1;
\frac{3}{2};\frac{-t_{1}}{t_2-t_1}\right)
\label{corr-theory}
\end{eqnarray}
with the hypergeometric function $_2F_1(z)$. The Brownian limit $\langle x(t)
x(t+\tau)\rangle=2D_0t$ follows for $\alpha=0$, while for $t_1=t_2$ we
recover Eq.~(\ref{MSD-theory}). The asymptotic behavior of expression
(\ref{corr-theory}), $\langle x(t)x(t+\tau)\rangle\simeq\tau^{(p-1)
/2}t^{(p+1)/2}$ for $\tau/t\gg1$ implies that the correlations decay with
$\tau$ for subdiffusive processes whereas they increase in the case of
superdiffusion, similar to fractional Brownian motion (FBM) \cite{deng}. We
also obtain the correlation of two consecutive increments along $x(t)$. For
$\tau/t\ll1$, we find the simple expressions for $p=1/2$ and $p=2$,
\begin{equation}
\label{increments-theory}
\langle[x(t)-x(t-\tau)][x(t+\tau)-x(t)]\rangle\sim\left\{\begin{array}{ll}
-\tau\sqrt{D_0/t}, & p=\frac{1}{2}\\(D_0\tau)^2, & p=2\end{array}\right..
\end{equation}
Indeed the occurrence of antipersistence (negative correlations) for
subdiffusion and persistence for superdiffusion holds for all values of $p$
\cite{supp}, again similar to FBM. We present exact results for the velocity
autocorrelation function in SM \cite{supp}.

To connect to single particle tracking experiments we now turn to the time
averaged MSD of a trajectory $x(t)$,
\begin{equation}
\label{TAMSD}
\left<\overline{\delta^2(\Delta)}\right>=\frac{1}{T-\Delta}\int\limits_{0}^{
T-\Delta}\left<\left[x(t+\Delta)-x(t)\right]^2\right>dt,
\end{equation}
where $\Delta$ is the lag time and $T$ the length of the time series $x(t)$. In
Eq.~(\ref{TAMSD}) we introduced the additional average over individual
trajectories, $\langle\overline{\delta^2(\Delta)}\rangle=N^{-1}\sum_{i=1}^N
\overline{\delta^2_i(\Delta)}$ \cite{he}. For $\Delta\ll T$ we find the linear
$\Delta$-dependence \cite{supp}
\begin{equation}
\label{TAMSD-theory}
\left<\overline{\delta^2(\Delta)}\right>=\frac{\Gamma(p+1/2)}{\pi^{1/2}}
\left(\frac{2}{p}\right)^{2p}D_0^p\frac{\Delta}{T^{1-p}}.
\end{equation}
Remarkably, this result holds for both sub- and superdiffusive regimes, and
we find the exact match $\langle\overline{\delta^2(\Delta)}\rangle=(\Delta/T)
^{1-p}\langle x^2(\Delta)\rangle$. Eq.~(\ref{TAMSD-theory}) is our central
result. Despite the simple nature of the Markovian HDP, we find that it
displays weak non-ergodicity, i.e., the scaling of
time and ensemble averages is different, $\langle\overline{\delta^2(\Delta)}
\rangle\neq\langle x^2(\Delta)\rangle$. Linear scaling of the type of
Eq.~(\ref{TAMSD-theory}) was found for subdiffusive CTRWs \cite{he}, but
also for correlated CTRWs \cite{corr}. It contrasts the ultraweak non-ergodicity
of superdiffusive CTRWs \cite{lws}. In particular, the dependence on the length
$T$ of the time series is identical to CTRW and correlated CTRW subdiffusion:
with increasing $T$ the effective diffusivity $D_{\mathrm{eff}}\simeq T^{p-1}$
decreases. In the superdiffusive regime, this $D_{\mathrm{eff}}$ increases with
time, the particle becomes more mobile. For HDP processes the ageing dependence
on $T$ arises when the particle continues to venture into regions of changing
diffusivities, for subdiffusive CTRWs this is due to increasingly longer
sojourn times.

Exploring HDPs in more detail numerically, we note that the Langevin equation
in the Stratonovich interpretation leads to an implicit mid-point iterative
scheme for the particle displacement. At step $i+1$, we thus have $x_{i+1}-x_i=
\sqrt{2D([x_{i+1}+x_i]/2)}(y_{i+1}-y_{i})$, where the increments $(y_{i+1}-y_i)$
of the Wiener process represent a $\delta$-correlated Gaussian noise with unit
variance. Unit time intervals separate consecutive iteration steps. In our
numerical study we consider the two generic cases $p=1/2$ and $p=2$. To avoid
divergencies in the discrete scheme, for subdiffusion we regularize the
diffusivity at $x=0$, choosing $D(x)=D_0A/(A+x^2)$, such that for $x^2\gg A$ we
recover the original relation (\ref{diff-coeff-power}). For superdiffusion, to
avoid trapping at $x=0$ and thus prohibitively expensive simulations, we use
the shifted form $D(x)=D_0(|x|+1)$, again with the correct scaling for large
$|x|$.

\begin{figure}
\includegraphics[width=8.8cm]{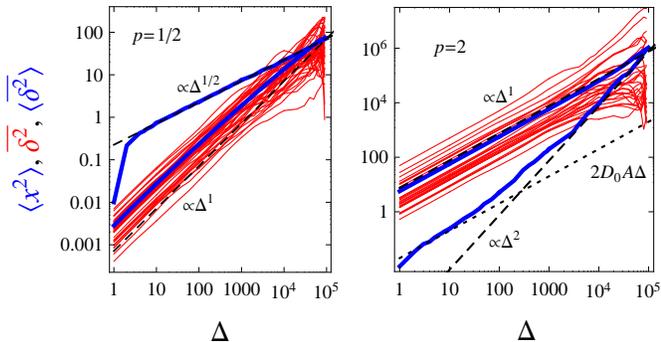}
\caption{Ensemble and time averaged MSDs for sub- and superdiffusive HDPs for
$T=10^5$, $D_0=1$, and $A=0.01$. Expected results (\ref{MSD-theory}) and (\ref{TAMSD-theory}) are shown
by dashed lines. The time averaged MSDs are logarithmically sampled. The
initial position is $x_0=0.1$ for all trajectories. For subdiffusion, the
discrepancy of theory and simulated results for $\langle\overline{\delta^2}
\rangle$ is due to the $D(x)$ rectification.}
\label{fig2}
\end{figure}

\emph{Numerical results.} From the generated trajectories, we compute the PDF
(Fig.~\ref{fig1}) as well as the ensemble and time averaged MSDs. The MSDs are
shown in Fig.~\ref{fig2}. For the ensemble averaged MSD (blue curve) we observe
excellent agreement with Eq.~(\ref{MSD-theory}). In the subdiffusion case, the
predicted behavior (\ref{MSD-theory}) is reached after less than a dozen steps,
such that the rectification of $D(x)$ and the choice of the off-center initial
position $x_0$ are practically negligible. For superdiffusion, an extended
region of almost normal diffusion is observed for small particle displacements,
as long as $|x|<1$ in the rectified $D(x)$. The long time behavior shows good
agreement between theory and simulations. The time averaged MSDs and their mean
$\langle\overline{\delta^2}\rangle$ follow nicely the predicted linear scaling.
Only when the lag time $\Delta$ approaches $T$ the linear behavior levels off
due to the finite trajectory length $T$.

Next we study the apparent diffusion coefficients and scaling exponents from
the individual $\overline{\delta^2}$. The points of each
trace $\overline{\delta^2}$ were logarithmically binned. The initial part of
the trajectory, $\Delta=1\ldots10^2$ steps, was fitted by a linear law with
diffusivity $D_1$. The long-time regime, $\Delta\sim10^2\ldots10^4$
steps, was fitted with two parameters, $\overline{\delta^2}=D_{\beta}
\Delta^{\beta}$. The distributions of $D_1$, $D_{\beta}$, and $\beta$ were then
fitted with the 3-parameter gamma distribution, $g(z)\sim z^{\nu-1}e^{-z/b}
e^{-a/z}$, the parameters being fixed by the first three moments obtained from
the histograms. This gamma distribution was recently proposed for the spread
with $\Delta$ of $\overline{\delta^2}$ of a Brownian walker \cite{grebenkov}.
In addition, we extracted the amplitude scatter PDF $\phi(\xi=\overline{\delta
^2}/\langle\overline{\delta^2}\rangle)$ of individual $\overline{\delta^2}$
around the mean $\langle\overline{\delta^2}\rangle$.

\begin{figure}
\includegraphics[width=8.8cm]{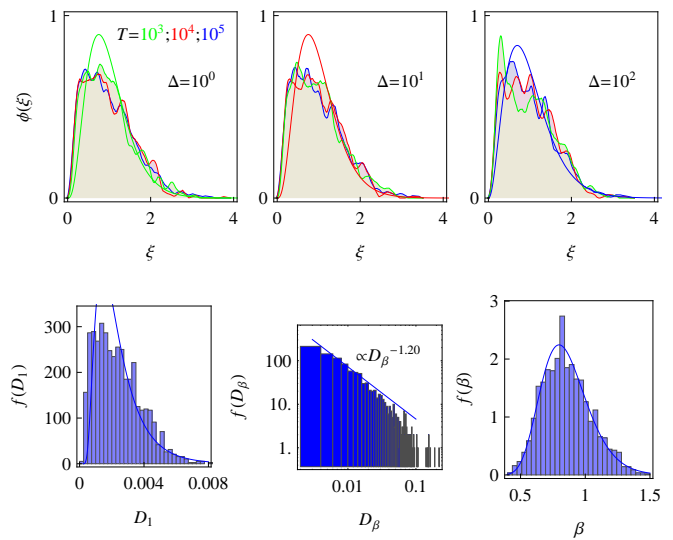}
\caption{Amplitude scatter PDF $\phi(\xi)$ for subdiffusve HDP (top) and
distributions of the fit parameters $D_1$, $D_{\beta}$, and $\beta$ (bottom).
Parameters $\Delta$ and $T$ are indicated in the panels. Solid envelopes
are 3-parameter fits to $g(z)$ (see text). At least $10^3$ trajectories were
analyzed for each $T$, with $T=10^5$.}
\label{fig3}
\end{figure}

\begin{figure}
\includegraphics[width=8.8cm]{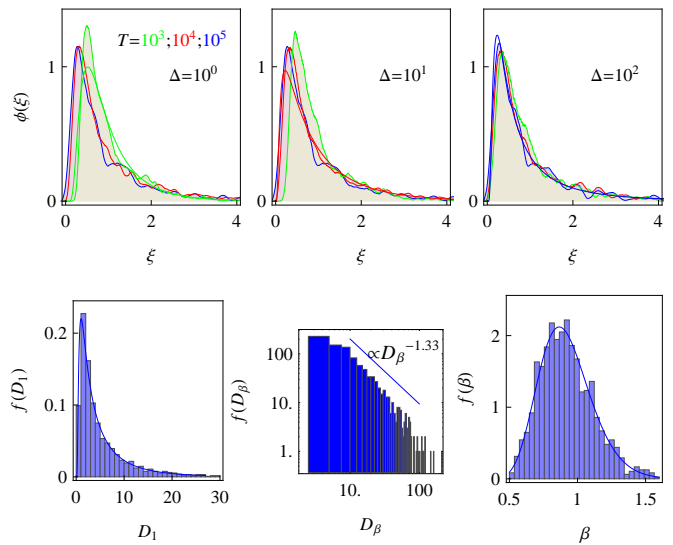}
\caption{The same as in Fig. \ref{fig3} but for superdiffusive HDPs.}
\label{fig4}
\end{figure}

Figs.~\ref{fig3} and \ref{fig4} show the results of this analysis, together
with the fits to the gamma distribution, $g(z)$. We see that the amplitude
scatter shows a pronounced spread around the ergodic value $\xi=1$ in both sub-
and superdiffusive regimes. The shape of $\phi(\xi)$ changes only marginally
with $\Delta$ in the linear region of $\overline{\delta^2}$. Broad
distributions of $\phi(\xi)$ are known from both sub- and superdiffusive CTRWs.
The differences are, however, that for subdiffusive CTRWs $\phi(\xi)$ has a
finite value at $\xi=0$ due to completely immobilized particles \cite{he}. For
superdiffusive CTRWs $\phi(\xi)$ decays to zero at $\xi=0$, but changes
significantly with $\Delta$ \cite{lws}. The scatter distribution $\phi(\xi)$
is thus a good criterion to distinguish HDPs and CTRWs.

More specifically, for subdiffusion the distributions of $D_1$ and the amplitude
scatter $\phi(\xi)$ are wider, while for superdiffusion they show a sharper peak
with a maximum shifted to the mean at $\xi=1$. The width of $\phi(\xi)$ is
roughly
unity for subdiffusion and $1/2$ for superdiffusion. It changes only slightly
with $T$ in the range $T=10^3\ldots10^5$, Fig.~\ref{fig3}. The distributions
for $D_\beta$ typically exhibit a long power-law tail with $f(D_{\beta})\sim
D_\beta^{-1.2\ldots1.4}$. The distributions for the apparent long-time exponent
$\beta$ are similar for sub- and superdiffusion, with $\langle\beta\rangle_{p=
1/2}\approx0.86$ and $\langle\beta\rangle_{p=2}\approx0.92$. Finite $T$-effects
lead to the undershoot of $\langle\beta\rangle$ compared to the theoretical
value of unity.

\begin{figure}
\includegraphics[width=8cm]{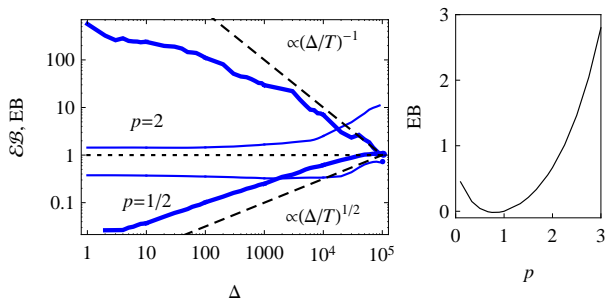}
\caption{Ergodicity breaking parameters $\mathcal{EB}$ (thick) and EB (thin
blue curves) as obtained from simulations ($T=10^5$). The theoretical
asymptotes for $\mathcal{EB}$ (left panel, dashed lines) and approximate
Eq.~(S9) for EB (right panel, see SM \cite{supp}).}
\label{fig5}
\end{figure}

Consider now the ergodicity breaking parameter \cite{he}
\begin{equation}
\label{EB1-HDP}
\mathrm{EB}=\lim_{T\to\infty}\frac{}{}
\left(\left<\left(\overline{\delta^2}\right)^2
\right>-\left<\overline{\delta^2}\right>^2\right)\Big/\left<\overline{\delta^2}
\right>^2
\end{equation}
that provides a good measure for the dispersion of time averaged MSDs for
different types of diffusion processes \cite{rytov}. The sufficient condition
of ergodicity of a process is $\mathrm{EB}=0$. A necessary condition is that
the ratio of the time and ensemble averaged MSD equals unity. Here we
observe that this condition is not fulfilled for $p\neq1$,
$\mathcal{EB}=\left<\overline{\delta^2(\Delta)}\right>\Big/\langle x^2(\Delta)
\rangle=(\Delta/T)^{1-p}$.
The ergodicity breaking parameters $\mathrm{EB}$ and $\mathcal{EB}$ extracted
from simulations data at $\Delta/T\ll1$ is $\mathrm{EB}\approx0.4$ for sub-
and $\mathrm{EB}\approx1.4$ for superdiffusion, indicating the
weakly non-ergodic nature of HDP processes in both regimes. The parameter
$\mathcal{EB}$ follows the predicted scaling for large
$\Delta$ and $T$ values, i.e., when both ensemble and time
averaged MSDs have converged to the theoretical scaling (Figs.~\ref{fig2}
and \ref{fig5}). For Brownian diffusion our simulations yield the correct
finite-time scaling $\mathrm{EB}(\Delta)=\frac{4}{3}\frac{\Delta}{T}$ and
sharp amplitude scatter $\phi(\xi)\sim\delta(\xi-1)$ at $T\to\infty$,
indicating the ergodic behavior for long traces and the self-averaging
property of normal diffusion. For $p=2$ at $\Delta/T\rightarrow0$ the
estimated theoretical value is $\mathrm{EB}=2/3$, (Eq.~(S9) in SM
\cite{supp}), close to the value from simulations.

The simulations show that for subfdiffusion, $\mathcal{EB}$ increases with
$\Delta/T$, while for superdiffusion it decreases with $\Delta/T$
(Fig.~\ref{fig5}). In both
cases $\mathcal{EB}$ approaches unity as $\Delta\to T$. At $\Delta/T\ll1$,
$\mathcal{EB}$ for $p<1$ assumes smaller values, indicating weaker deviations
from ergodicity, in contrast to superdiffusion, see Fig.~\ref{fig5}. The
parameter $\mathrm{EB}$ defined via the fourth moment is overall a more robust
characteristic of the process. For instance, it varies only slightly with $T$
and $\Delta$ for $\Delta/T\ll1$ (not shown).

\emph{Conclusions.} The seemingly simple Markovian HDP with power-law
space-dependent
diffusion coefficient is defined in terms of a Langevin equation, which is
fully local in both space and time. Despite this locality HDPs are weakly
non-ergodic: the time averaged MSD $\overline{\delta^2 (\Delta)}$ scales
linearly with the lag time $\Delta$ for both sub- and superdiffusion. Such
a behavior was previously known for subdiffusion from more complex processes:
CTRWs with diverging characteristic waiting time \cite{he} and CTRWs with
correlated waiting times \cite{corr}. For superdiffusion only an ultraweak
ergodicity breaking was reported in which the coefficients of the time and
ensemble MSDs differ \cite{lws}, while space-correlated CTRWs feature a
quadratic scaling of the time averaged MSD \cite{corr}.

The amplitudes of $\overline{\delta^2 (\Delta)}$ of different trajectories
show a broad and asymmetric distribution, another sign of non-ergodicity. In
contrast to CTRW subdiffusion, however, in HDPs there is no contribution from
completely immobile trajectories.
Concurrently, HDPs exhibit (anti)persistence of the increment correlation
functions, such that for subdiffusion (superdiffusion) subsequent increments
preferentially have opposite (equal) direction. This property is similar to
the noise correlator of FBM, moreover the velocity autocorrelator of both
processes is strikingly similar.

HDPs represent new tools for the description of weakly non-ergodic dynamics.
Due to their intuitive formulation in terms of space-dependent diffusivities
this dynamic behavior of HDPs directly follows from the physical properties
of the environment. HDPs may be distinguished from other processes due to
the difference in the amplitude scatter distribution of time averaged MSDs,
the increment autocorrelation function, as well as the two ergodicity breaking
parameters. At the same time our observations for HDPs pose the interesting
question for the minimal necessary condition for the occurrence of weak
ergodicity breaking.

The authors thank E. Barkai and H. Buening for discussions and acknowledge
funding from the Academy of Finland (FiDiPro scheme to RM) and Deutsche
Forschungsgemeinschaft (Grant CH 707/5-1 to AGC).


\begin{thebibliography}{99}

\bibitem{perrin} J. Perrin, Comptes Rendus \textbf{146}, 967 (1908); I.
Nordlund, Z. Phys. Chem. \textbf{87}, 40 (1914).

\bibitem{spt} C. Br{\"a}uchle, D. C. Lamb, and J. Michaelis, Single Particle
Tracking and Single Molecule Energy Transfer (Wiley-VCH, Weinheim, 2010);
X. S. Xie et al., Annu. Rev. Biophys.
\textbf{37}, 417 (2008).

\bibitem{ran} R. Nathan et al., Proc. Natl. Acad. Sci. USA \textbf{105},
19052 (2008); D. W. Sims et al., Nature \textbf{451}, 1098 (2008).

\bibitem{dirk} M. C. Gonz{\'a}lez, C. A. Hidalgo, and A.-L. Barab{\'a}si,
Nature \textbf{453}, 779 (2008); D. Brockmann, L. Hufnagel, and T. Geisel,
\emph{ibid.} \textbf{439}, 462 (2006).

\bibitem{report} R. Metzler and J. Klafter, Phys. Rep. \textbf{339}, 1 (2000);
J. Phys. A \textbf{37}, R161 (2004).

\bibitem{weber} I. Golding and E. C. Cox, Phys. Rev. Lett. \textbf{96}
098102 (2006); S. C. Weber, A. J. Spakowitz, and J. A. Theriot, Phys. Rev.
Lett. \textbf{104}, 238102 (2010); I. Bronstein et al., Phys. Rev. Lett.
\textbf{103}, 018102 (2009).

\bibitem{krapf} A. V. Weigel, B. Simon, M. M. Tamkun, and D. Krapf, Proc. Nat.
Acad. Sci. USA \textbf{108}, 6438 (2011).

\bibitem{lene} J.-H. Jeon, V. Tejedor, S. Burov, E. Barkai, C. Selhuber-Unkel,
K. Berg-S{\o}rensen, L. Oddershede, and R. Metzler, Phys. Rev. Lett.
\textbf{106}, 048103 (2011).

\bibitem{seisenhuber} G. Seisenberger et al., Science \textbf{294}, 1929 (2001).

\bibitem{elbaum} A. Caspi, R. Granek, and M. Elbaum, Phys. Rev. Lett.
\textbf{85}, 5655 (2000); Phys. Rev. E \textbf{66}, 011916 (2002).

\bibitem{rogers} S. S. Rogers, C. v. d. Walle, and T. A. Waigh, Langmuir
\textbf{24}, 13459 (2008).

\bibitem{mussels} M. de Jager, F. J. Weissing, P. M. J. Herman,
B. A. Nolet, and J. van de Koppel, Science \textbf{332}, 1551 (2011).

\bibitem{aphids} A. Mashanova, T. H. Oliver, V. A. A. Jansen, J. R. Soc.
Interface \textbf{7}, 199 (2010).

\bibitem{pt} E. Barkai, Y. Garini, and R. Metzler, Phys. Today \textbf{65}(8),
29 (2012).

\bibitem{igor} I. M. Sokolov, Soft Matter \textbf{8}, 9043 (2012).

\bibitem{yazmin} Y. Meroz, I. Eliazar, and J. Klafter, J. Phys. A \textbf{42},
434012 (2009); Y. Meroz, I. M. Sokolov, and J. Klafter, Phys. Rev. Lett.
\textbf{110}, 090601 (2013).

\bibitem{deng} W. Deng and E. Barkai, Phys. Rev. E \textbf{79}, 011112 (2009).

\bibitem{jae} J.-H. Jeon and R. Metzler, Phys. Rev. E \textbf{85}, 021147 (2012).

\bibitem{goychuk} I. Goychuk, Phys. Rev. E \textbf{80}, 046125 (2009); Adv.
Chem. Phys. \textbf{150}, 187 (2012).

\bibitem{montroll} E. W. Montroll and G. H. Weiss, J. Math. Phys. \textbf{6},
167 (1965); H. Scher and E. W. Montroll, Phys. Rev. B \textbf{12}, 2455
(1975).

\bibitem{web} J.-P. Bouchaud, J. Phys  I \textbf{2}, 1705 (1992); G. Bel and
E. Barkai, Phys. Rev. Lett. \textbf{94}, 240602 (2005); A. Rebenshtok and E.
Barkai, \emph{ibid.} \textbf{99}, 210601 (2007).

\bibitem{he} Y. He, S. Burov, R. Metzler and E. Barkai, Phys. Rev. Lett.
\textbf{101}, 058101 (2008); A. Lubelski, I. M. Sokolov, and J. Klafter,
\emph{ibid.} \textbf{100}, 250602 (2008); S. Burov, J.-H. Jeon, R. Metzler,
and E. Barkai, Phys. Chem. Chem. Phys. \textbf{13}, 1800 (2011).

\bibitem{eli} F. D. Stephani, J. P. Hoogenboom, and E. Barkai, Phys. Today
\textbf{62}(2), 34 (2009).

\bibitem{richardson} L. F. Richardson, Proc. Roy. Soc. London Ser. A
\textbf{110}, 709 (1926).

\bibitem{dentz} R. Haggerty and S. M. Gorelick, Water Res. Res. \textbf{31},
2383 (1995); M. Dentz et al., Adv. Water Res. \textbf{49}, 13 (2012).

\bibitem{loverdo} C. Loverdo et al., Phys. Rev. Lett. \textbf{102}, 188101
(2009); B. O'Shaughnessy and I. Procaccia, Phys. Rev. Lett.
\textbf{54}, 455 (1985).

\bibitem{kuehn} B. P. English, V. Hauryliuk, A. Sanamrad, S. Tankov, N. H.
Dekker,, and J. Elf, Proc. Natl. Acad. Sci. \textbf{108}, E365 (2011);
T. Kuehn et al., PLoS One \textbf{6}, e22962 (2011).

\bibitem{platani} M. Platani, I. Goldberg, A. I. Lamond, and J. R. Swedlow,
Nature Cell Biol. \textbf{4}, 502 (2002).

\bibitem{risken} H. Risken, The Fokker-Planck Equation, (Springer, Heidelberg,
1989).

\bibitem{REM} See the explanation in B. J. West, A. R. Bulsara, K. Lindenberg,
V.  Seshadri, and K. E. Shuler, Physica A \textbf{97}, 211 (1979). The
corresponding diffusion equation for the PDF $P(x,t)$ has the symmetrized form
\[
\frac{\partial P(x,t)}{\partial t}=\frac{\partial}{\partial x}\left[\sqrt{D(
x)}\frac{\partial}{\partial x}\left(\sqrt{D(x)}P(x,t)\right)\right].
\]

\bibitem{footnote2} In our model $\alpha<2$ to ensure the growth condition
for existence and uniqueness of the solution of a Markovian stochastic
differential equation, see I. I. Gikhman, A. V. Skorokhod, Stochastic
Differential Equations (Springer, Heidelberg, 1972) and C. W. Gardiner,
Handbook of Stochastic Methods (Springer, Heidelberg, 2004).

\bibitem{supp} Supplementary Material.

\bibitem{abm} R. Metzler and J. Klafter, Europhys. Lett. \textbf{51}, 492 (2000);
R. Metzler and T. F. Nonnenmacher, Phys. Rev. E \textbf{57} 6409 (1998).

\bibitem{corr} M. Magdziarz, R. Metzler, W. Szczotka, and P. Zebrowski,
Phys. Rev. E \textbf{85}, 051103 (2012); V. Tejedor and R. Metzler, J.
Phys. A \textbf{43}, 082002 (2010).

\bibitem{lws} A. Godec and R. Metzler, Phys. Rev. Lett. \textbf{110}, 020603
(2013); D. Froemberg and E. Barkai, E-print arXiv:1211.1539.

\bibitem{grebenkov} A. Andreanov and D. S. Grebenkov, J. Stat. Mech.,
p07001 (2012).

\bibitem{rytov} S. M. Rytov, Yu. A. Kravtsov, and V. I. Tatarskii, Principles
of Statistical Radiophysics 1: Elements of Random Process Theory (Springer,
Heidelberg 1987).

\end{thebibliography}
\end{document}